\documentstyle[prl,aps,twocolumn]{revtex}
\draft
\tightenlines
\newcommand{\be}{\begin{equation}}
\newcommand{\ee}{\end{equation}}
\newcommand{\bea}{\begin{eqnarray}}
\newcommand{\eea}{\end{eqnarray}}
\begin{document}
\title {An Exactly Solvable Many-Body Problem in One Dimension}
\author {Sudhir R. Jain$^*$}
\address {Theoretical Physics Division \\
Bhabha Atomic Research Centre,\\
Trombay, Mumbai 400 085, India}
\vskip .5 true cm
\author {Avinash Khare$^+$}
\address {Institute of Physics, Sachivalaya Marg\\
 Bhubaneswar 751 005, Orissa, India}
\maketitle

\begin{abstract}

For $N$ impenetrable particles in one dimension where only the nearest and
the next-to-nearest neighbours
interact, we obtain the complete spectrum both on a line
and on a circle.
Further, we establish a mapping
between these $N$-body problems and the short-range
Dyson model introduced recently to model intermediate spectral
statistics in some systems using which we compute the two-point
correlation function and prove the absence of long-
range order in the corresponding many-body theory.
Further, we also show the absence of an off-diagonal
long-range order in the system.
\vskip 0.25 truecm

\end{abstract}
\noindent
PACS Nos.  05.50.+q, 05.90.+m, 03.65.Ge, 05.30.-d, 05.45.+b
\vskip 0.25 truecm

Connections between exactly solvable models \cite{mattis}
and random matrix theory \cite{mehta}
have been very fruitful. The Calogero-Sutherland models \cite{calogero}
are some of the well-known examples which are not only interesting
because they are exactly solvable \cite{olsha} but also due to their
relationship with (1+1)-dimensional conformal field theory, matrix models
etc. \cite{simons}. Recent developments \cite{srednicki,our}
relating equilibrium
statistical mechanics to random matrix theory owing to non-integrability
of dynamical systems has made the pursuit of unifying seemingly
disparate ideas a very important theme.

The family consisting of exactly solvable models, related to fully integrable systems,
is quite small \cite{op} and their importance lies in the fact that their small
perturbations describe wide range of physically interesting situations.

The universality in level correlations in linear (Gaussian) random matrix
ensembles
agrees very well with those in chaotic quantum systems \cite{bohigas}
as also in many-body systems like nuclei \cite{mehta}. On the other hand, random
matrix theory was connected to the world of exactly solvable models when the
Brownian motion model was presented by Dyson \cite{mehta},
and later on, by the works on level dynamics \cite{haake}. However, there
are dynamical systems which are neither chaotic nor integrable - the so-called
pseudointegrable systems \cite{jain-lawande}.
It is known that the spectral statistics of such
systems are ``non-universal with a universal trend" \cite{parab-jain}.
In particular, for Aharonov-Bohm billiards, the level spacing distribution
is linear for small spacing and it falls exponentially for large spacing
\cite{murthy}. Similar features are numerically observed for the Anderson model
in three dimensions at the metal-insulator transition point \cite{guhr}.
To understand these statistical features, and in the context of random banded
matrices, a new random matrix model (which has been called as the short-range
Dyson model in \cite{gremaud}) was introduced \cite{pandey,bogomolny} wherein
the energy levels are treated as in the Coulomb gas model with the difference
that only nearest neighbours interact. This new model explains features
of intermediate statistics \cite{gremaud} in some polygonal billiards.

In view of all this it is worth enquiring if one can construct an $N$-body
problem which is exactly solvable and which is connected to the short-range
Dyson model (SRDM)?
The purpose of
this letter is to present such a model.
We find the density and the two-point correlation
function of the correponding many-body theory from where
we conclude that there is no long-range order in the thermodynamic limit.
Further, we also prove  the
absence of off-diagonal long-range order (ODLRO) in the system.

Let us first discuss the N-body problem on a line.
We begin by writing the Hamiltonian for the $N$-body problem
($\hbar = m =1)$ :
\bea
H &=& - {1\over 2} \sum^N_{i=1} {\partial^2\over \partial x^2_i} +
g \sum^{N-1}_{i=1}
{1\over (x_i-x_{i+1})^2}\nonumber \\ &-& G \sum^{N-1}_{i=2}
{1\over (x_{i-1}-x_i)(x_i-x_{i+1})}
+ V(\sum^N_{i=1} x^2_i).
\eea
We shall show that the complete bound state spectrum of $H$ can be obtained if
\be\label{2}
g = \beta (\beta-1), \ G = \beta^2,
V = {\omega^2\over 2} \sum^N_{i=1} x_i^2 \, .
\ee
In (1), $g$ and $G$ are the strengths of two-body and three-body potentials
respectively.
Using the ansatz,
\be
\psi = \phi \bigg [\prod^{N-1}_{i=1} (x_i-x_{i+1})^{\beta} \bigg ] \, ,
\ee
in the corresponding Schr\"odinger equation $H\psi = E\psi$,
$\phi$ is seen to satisfy the equation
\bea
- {1\over 2} \sum^N_{i=1} {\partial^2\phi\over \partial x^2_i} &-& \beta
\sum^{N-1}_{i=1} {1\over (x_i-x_{i+1})} \left({\partial \phi\over\partial x_i}
-{\partial
\phi\over \partial x_{i+1}}\right)\nonumber \\
&+&(V-E)\phi = 0 \, .
\eea
On substituting
\be
\phi = P_k (x) \Phi (r), \ \  r^2 \equiv \sum^N_{i=1} x^2_i \, ,
\ee
$\Phi (r)$ satisfies the equation
\bea
\Phi^{''} (r) &+& [N +2k-1+2(N-1)\beta]{1\over r}\Phi' (r) \nonumber \\
&+& 2\left[ E-V(r)\right] \Phi (r) = 0 \, ,
\eea
and  $P_k (x)$, a translation-invariant homogeneous polynomial of degree
$k$ ($k = 0,1,2,...$) in the particle-coordinates satisfies generalised
Laplace equation,
\bea
&& \bigg [ \sum^N_{i=1} {\partial^2 \over \partial x^2_i} + 2\beta
 \sum^{N-1}_{i=1} {1\over (x_i-x_{i+1})} \nonumber \\
&& \left({\partial \over\partial x_i}
-{\partial
\over \partial x_{i+1}}\right) \bigg ] P_k(x) = 0 \, .
\eea
Specializing to the case of the oscillator potential i.e. $V(r)
= {\omega^2\over 2} r^2$, we see that
\be
\Phi (r) = \exp ({-\omega r^2/2}) L^a_n (\omega r^2), \ n = 0,1,2,....
\ee
where $L^{a}_n(x)$ is the associated Laguerre polynomial while the energy
eigenvalues are given by
\bea\label{9}
E_{n,k} &=& \left[2n+k+{N\over 2} + (N-1)\beta \right] \omega \nonumber \\
&=& E_0 + (2n+k) \omega \, ,
\eea
with $a = {E\over \omega}-2n-1$.
Note that
$\lim_{N\rightarrow\infty} {E\over N} = \left(\beta +{1\over 2}\right)\omega $,
thus,  the system has a good thermodynamic limit. In contrast, notice that the
long-ranged Calogero model does not have good thermodynamic limit since in that
case for large $N$, $E/N$ goes like $N$ (and is not a constant).
The spectrum (\ref{9}) can be interpreted as due to noninteracting bosons
(or fermions) plus
$(n,k)$-independent (but $N$-dependent) shift. It is instructive to note that
as in the Calogero case \cite{ak}, the
entire discrete spectrum of the $N$-body problem can also be obtained if
$V(\sum_i x_i^2) = -{\alpha \over \sqrt{\sum_i x_i^2}}$.

Thus, the ground state solution  is  given by
($n = k = 0$)
\be
\psi_0 = \exp \left({-{\omega\over 2}\sum^N_{i=1}x^2_i}\right) \prod^{N-1}_{i=1}
(x_i-x_{i+1})^{\beta} \, ,
\ee
\be
E_0 = \left[ (N-1)\beta+{N\over 2}\right] \omega \, .
\ee
The probability distribution for $N$ particles is given by
\be
\psi^{2}_0 =
\mbox{constant~~} \exp \left( -\beta \sum^N_{i=1} y^2_i\right) \prod^{N-1}_{i=1} (y_i-y_{i+1})^{2\beta}
\ee
where
$y_i \equiv \sqrt{{\omega\over\beta}} x_i $.
We observe that $\psi^{2}_0$ is the joint probability distribution
function for eigenvalues in the SRDM.
Using this mapping one can now calculate the correlation functions of the
many-body theory defined in the limit $N \rightarrow \infty, \omega
\rightarrow 0, N\omega$  fixed. For example,
the one-point function (density) tends to a Gaussian for any $\beta $
\cite{pandey} and is given by
\be
R_1(x) = \frac{N}{\sqrt{2\pi \sigma ^2}}\exp
\left( -\frac{x^2}{2\sigma ^2}\right),
\ee
where $\sigma ^2 = \frac{(\beta + 1)}{\omega}$.
Before we discuss the higher correlation functions
we first consider another $N$-body problem
where we impose periodic boundary conditions on
the wavefunction and obtain the ground state energy as well
as the excitation spectrum; since in the thermodynamic limit the
correlation functions of the two are the same.
The Hamiltonian is
\bea
H &=& - {1\over 2} \sum^N_{i=1}{\partial^2\over\partial x^2_i}
+ g{\pi^2\over L^2} \sum^{N-1}_{i=1}
{1\over \sin^2 [{\pi\over 2}(x_i-x_{i+1})]} \nonumber \\
&-& G \sum^{N-1}_{i=2} \cot \left[ (x_{i-1}
-x_i) {\pi\over L}\right] \cot \left[(x_i-x_{i+1}){\pi\over L} \right].
\eea
First, we wish to find the ground state subjected to the periodic boundary
condition
\be
\psi(x_1,...,x_i+L, ..., x_N) = \psi (x_1,..., x_i,..., x_N).
\ee
To this end, we start with a trial wavefunction of the form
\be
\Psi_0 = \prod^{N-1}_{i=1} \sin^{\beta} \left[{\pi\over L} (x_i-x_{i+1})\right]
\ee
and substitute it in
the Schr\"odinger equation. We find that
it is indeed a solution provided $g$ and $G$ are again as given by
eq. (\ref{2}).
The corresponding energy eigenvalue turns out to be
\be
E_0 = {N \over L^2} \beta^2\pi^2 \, .
\ee
Thus, unlike the Calogero-Sutherland type of models, our models
 have good thermodynamic limit i.e. the ground
state energy per particle $(=E_0/N)$ is finite.

The excitation spectrum can also be obtained in this case the details of which
will be given elsewhere \cite{jk} and it can be shown that
\be
E = E_0 +{2\pi^2 \over L^2} [\sum_{j} n_j^2 +\beta (\sum n_1 -1)]
\ee
where $\sum n_1$ means total number of quanta in the state ``1'' when one
includes all possible configurations for a given $N$. Here $n_j (j = 1,2,...
,N$) are ordered according to $n_1 \le n_2 \le ...\le n_N$.

Having obtained the exact ground state, it is natural to enquire if
the corresponding $N$-particle probability density can be mapped to the joint
probability distribution of some short-range circular Dyson model (SRCDM) so
that we can obtain some exact results for the corresponding many-body theory.
Indeed the square of the ground-state wavefunction is related to the
joint probability distribution function for the SRCDM. The density
is a constant if $0 < x < N/L$, and is zero outside. The two-point
level correlation function is the same in the SRDM as well as SRCDM.
In fact Pandey \cite{pandey} has already evaluated it. In particular, on
concentrating on the circular version of the many-body problem for simplicity
he has shown that \cite{pandey}
\be
R_2^{(\beta )}(s) = \sum_{n=1}^{\infty } P^{(\beta )}(n,s),
\ee
where
\be
P^{(\beta )}(n,s) = \frac{(\beta +1)^{n(\beta +1)}}{\Gamma [n(\beta +1)]}
s^{n(\beta +1)-1}e^{-(\beta +1)s}.
\ee
On summing the infinite series, we obtain the following result
for any integer $\beta $ :
\be\label{20}
R_2^{(\beta )}(s) = \sum_{k=0}^{\beta } \Omega ^k
e^{(\beta +1)s\{\Omega ^k-1\}}
\ee
where $\Omega = e^{2\pi i/(\beta +1)}$. For $\beta =1$,
which corresponds to orthogonal ensemble, the result is already known
\cite{bogomolny,pandey}, $R_2 (\beta=1,s) = 1-e^{-4s}$.
Using eq. (\ref{20}) we can also evaluate it for other integral values of
$\beta$. In particular, for $\beta =2,4,3$ (the first two of which correspond
to unitary and symplectic ensembles respectively) we find that
\be
R_2^{(2)}(s) = 1 - 2e^{\frac{-9s}{2}}\cos \left(\frac{3\sqrt{3}s}{2}
- \frac{\pi}{3}\right);
\ee
\bea
R_2^{(4)}(s) &=& 1 + 2e^{5s(-1+\cos (2\pi /5))}\cos \left[{2\pi \over 5}+5s\sin \left({2\pi \over 5} \right) \right] \nonumber \\
&+& 2e^{5s(-1+\cos (4\pi /5))}\cos \left[{4\pi \over 5}+5s\sin \left({4\pi \over 5} \right) \right];
\eea
\be
R_2^{(3)}(s) = 1 - e^{-8s} - 2e^{-4s}\sin (4s).
\ee
In Fig. 1 we have plotted $R_2 (s)$ as a function of $s$ for $\beta =
1,2,3,4$.
From the figure as well as the general result given above
we conclude that
 there is no long-range
order in the corresponding many-body theory.
For other values of $\beta$, we have an integral representation which we
have not succeeded to bring to a closed form.

We now prove that there is no off-diagonal
long-range order either in this many-body theory.
For the proof, we again concentrate on the circular version of the many-body
problem for
simplicity. Precisely, we shall prove that the Penrose-Onsager criterion
\cite{penrose-onsager} is fulfilled , i.e., the largest eigenvalue of the
density matrix, divided by the total number of particles tends to
zero in the thermodynamic limit.

Denoting the wavefunction in (16) by $\psi _{N,L}(x_1,...,x_N)$, the
one-particle reduced density matrix  and its
Fourier coefficients (all non-negative) are given by ($\xi = x-x'$)
\bea
\rho _{N,L}(\xi ) &=& N\int_{0}^{L}dx_1...\int_{0}^{L}dx_{N-1}
\psi _{N,L}(x_1,...,x_{N-1},x) \nonumber \\
&&\psi _{N,L}^{\star}(x_1,...,x_{N-1},x'),
\eea
\be
\rho _{N,L}^{(n)} = \int_{0}^{L} \rho _{N,L}(\xi )e^{i2\pi n\frac{\xi }{L} }d\xi .
\ee
This, in turn, is connected to the momentum distribution function,
\be
F_{N,L}(k) = \frac{1}{N} \sum_{-\infty < n < \left(\frac{kL}{2\pi}\right)}
\rho _{N,L}^{(n)}.
\ee
Equivalently, we can also write
\be
\frac{L}{N}\rho _{N,L}(\xi ) = \int_{-\infty }^{\infty }e^{ik\xi }dF_{N,L}(k).
\ee
Since the box size enters throughout only as a scale factor, we write
$\rho _{N,L}(\xi ) = \frac{1}{L}R_N(\alpha ), ~~\alpha = \frac{2\pi \xi}{L}$.
After some manipulations, $R_N(\alpha )$ can  be written as
\bea
R_N(\alpha ) &=& \frac{2^{2(N-3)+4\beta }}{[(N-1)!]^2(2\pi )^{N-1}}
\int_{0}^{2\pi}d\theta _1...\int_{0}^{2\pi}d\theta _{N-2}
\nonumber \\
&& \prod_{n=1}^{N-3}
\sin ^{2\beta} (\frac{\theta _{n}-\theta _{n+1}}{2}){\cal I}(\alpha
, \theta _{N-2}) \, ,
\eea
where
\bea
{\cal I}(\alpha , \theta _{N-2})
&=& \int_{0}^{2\pi}d\theta _{N-1}
\sin ^{2\beta}\left(\frac{\theta _{N-2}-\theta _{N-1}}{2} \right)
\nonumber \\
&& \sin ^{2\beta}\left(\frac{\theta _{N-1}-\alpha}{2}\right) \, .
\eea
Because of positivity of
$\rho _{N,L}(\xi )$, the largest discontinuity of $F_{N,L}(k)$ occurs at
the origin and that is \cite{lenard}
\bea
\frac{\rho _{N,L}^{(0)}}{N} &=& \frac{1}{2\pi N}\int_{-\pi}^{\pi}
R_N(\alpha )d\alpha \nonumber \\
&<& \frac{2^{2(N-3)+4\beta}}{N!(N-1)!} \, .
\eea
The inequality is easily shown by estimating the integral and
since this ratio tends to zero as $N\rightarrow \infty$, we have proved
that the Penrose-Onsager criterion is fulfilled for the absence of
Bose-Einstein condensation. In fact, we have shown above that the
one-particle reduced density matrix does not obtain ODLRO\cite{foot}.

In this context it is worth pointing out that in the Calogero-type model
in two dimensions, recently one of us had shown the existence of ODLRO \cite{kr}.
Finally, we have been able to generalize some of these results to
higher dimensions as well as to other root systems, the details of which will
be published elsewhere \cite{jk}.

To conclude, (i) we have exactly solved a quantum many-body problem
with nearest and next-to nearest neighbour interactions
in one dimension;
(ii) we have shown the relation between these many-body problems and
the short-range Dyson model, developed recently
to understand intermediate spectral statistics in pseudointegrable systems
and the 3-D Anderson model at the metal-insulator transition point;
(iii) it is shown that
there is no long-range order in the many-body system; and, (iv) it
is proven that off-diagonal long-range order is absent, so no
quantum phase transitions can occur.
The model presented here and its complete treatment puts it in the class of
celebrated Toda lattices \cite{toda} where the interaction is among
nearest neighbours.
\vskip 0.25 truecm
\noindent
{\bf Acknowledgement}   One of us (SRJ) acknowledges the warm hospitality of
the Institute of Physics, Bhubaneswar where this work was initiated.
We are also grateful to Akhilesh Pandey for making his unpublished work
available to us, and to Zafar Ahmed and
M. A. Prasad for their invaluable help in summing the series for the
two-point function.
\vskip 0.05 truecm
\noindent
{\bf Figure Caption}
\noindent
Fig. 1  The two-point correlation function for four integer values of $\beta $
(small dashes to larger dashes denote from $\beta$ varying from 1 to 4)
clearly shows an absence of long-range order.
\vskip 0.05 truecm
\noindent
* srjain@apsara.barc.ernet.in

\noindent
+ khare@iopb.res.in


\begin{references}
\bibitem{foot} It is also possible to show that the two-particle
reduced density matrix does not obtain ODLRO.
\end{references}
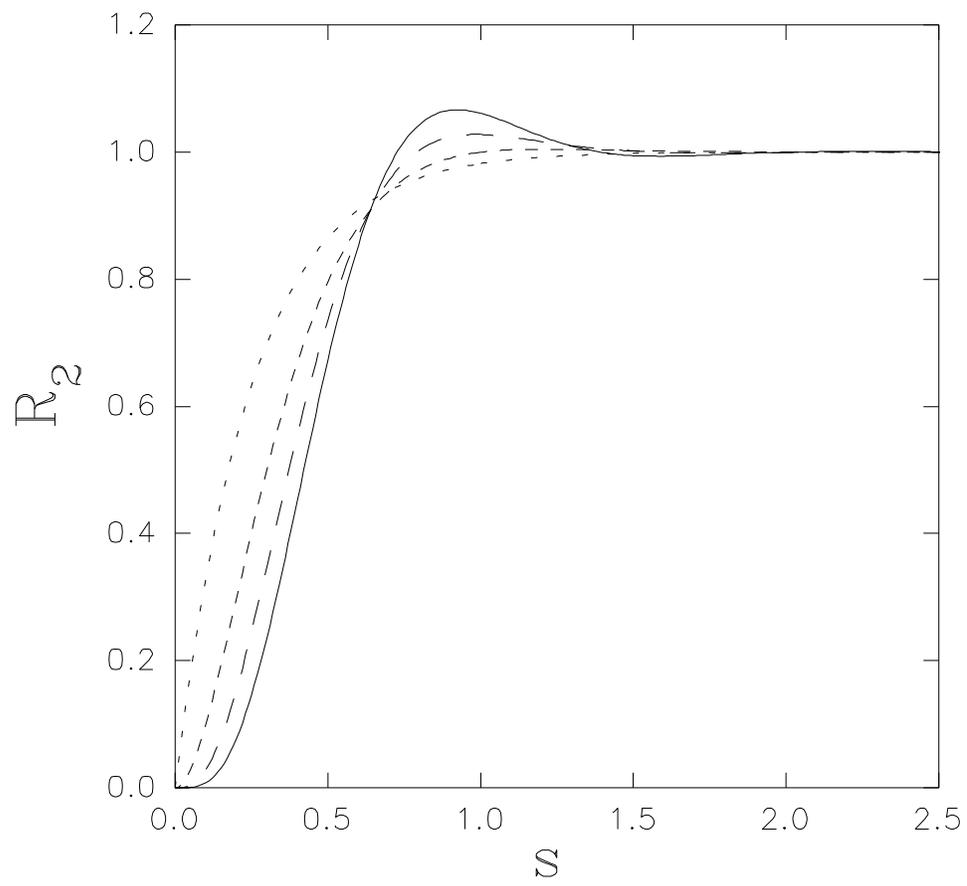
\end{document}